# Electronic structure and chemical bonding in novel tetragonal phase $Ca_{10}(Pt_4As_8)(Fe_2As_2)_5$ as a parent material for the new family of high-$T_C$ iron-pnictide superconductors.

I.R. Shein,* A.L. Ivanovskii

*Institute of Solid State Chemistry, Ural Branch of the Russian Academy of Sciences, 620990 Ekaterinburg, Russia*

A B S T R A C T

By means of first-principles calculations, the electronic structure and chemical bonding for the recently discovered tetragonal (s.g. *P4/n*; # 85) superconducting ($T_C \sim$ 25K) phase $Ca_{10}(Pt_4As_8)(Fe_2As_2)_5$ have been examined in details, and the optimized structural parameters, electronic bands, densities of states, and inter-atomic bonding picture were evaluated and analyzed in comparison with related layered iron-based superconducting materials. We have shown that (i). $Ca_{10}(Pt_4As_8)(Fe_2As_2)_5$ is metallic-like, and the electronic bands in the window around the Fermi level are formed mainly by the Fe $3d$ states of $(Fe_2As_2)_5$ blocks; (ii). the $(Pt_4As_8)$ blocks will behave as semi-metals with very low densities of states at the Fermi level; (iii). the near-Fermi bands adopt a "mixed" character: simultaneously with quasi-flat bands, a series of high-dispersive bands which intersect the Fermi level was found; (iv). the of chemical bonding in $Ca_{10}(Pt_4As_8)(Fe_2As_2)_5$ is very complicated and includes an anisotropic mixture of covalent, metallic, and ionic inter-atomic and inter-block interactions.

* Corresponding author. *E-mail address:* shein@ihim.uran.ru (I.R. Shein).

# 1. Introduction

Since the discovery of superconductivity in the layered LaFeAsO$_{1-x}$F$_x$ with $T_C \sim 26$K [1], intensive research efforts have focused on search of new Fe-based superconductors (SCs). Today a lot of related iron-pnictide (Fe-$Pn$) and iron-chalcogenide (Fe-$Ch$) SCs have been successfully developed. These materials are united into several groups, namely so-called 11 (s.g. $P4/nmm$, such as FeSe), 111 (s.g. $P4/nmm$, such as LiFeAs), 122 (s.g. $I4/mmm$, such as $A$Fe$_2$$Pn_2$, or $A$Fe$_2$Se$_2$, $A$ = K, Ca, Sr, Ba), and 1111 - like (s.g. $P4/nmm$, such as $RE$Fe$Pn$O, $RE$ = rare earths) systems, reviews [2-8]. Besides, recently this family was even more expanded, and as parent phases for the new high-$T_C$ SCs, more complex compounds such as (Fe$_2$$Pn_2$)(Sr$_4$$M_2$O$_6$), where $M$ are 3$d$ metals (review [9]), as well as (Fe$_2$As$_2$)(Ca$_{n+2}$(Al,Ti)$_n$O$_y$) (n = 2, 3, 4) [10], (Fe$_2$(P,As)$_2$)(Ca$_4$Al$_2$O$_{6-y}$) [11], (Fe$_2$As$_2$)(Sr$_4$(Sc,Ti)$_3$O$_8$), (Fe$_2$As$_2$)(Ba$_4$Sc$_3$O$_{7.5}$), (Fe$_2$As$_2$)(Ba$_3$Sc$_2$O$_5$) [12], (Fe$_2$As$_2$)(Ca$_4$(Mg,Ti)$_3$O$_y$) [13], (Fe$_2$$Pn_2$)(Sr$_4$MgTiO$_6$) [14,15], and (Fe$_2$As$_2$)(Ba$_4$Sc$_2$O$_6$) [16] were prepared, demonstrating high chemical and structural flexibility of these materials.

Note that the structures of all known iron-pnictide and iron-chalcogenide materials include quasi-two-dimensional (2D-like) (Fe$_2$$Pn_2$) (or (Fe$_2$$Ch_2$)) blocks, which are separated by $A$ atomic sheets or ($RE$O) blocks (for 111, 122, and 1111-like phases, respectively) or by perovskite-like [$A_3$$M_2$O$_5$], [$A_2$$M$O$_3$], [Ca$_{n+1}$(Sc,Ti)$_n$O$_y$] $etc.$ oxide blocks for the aforementioned more complex phases. Besides, the electronic bands in the window around the Fermi level, $E_F$, are formed mainly by the states of these (Fe$_2$$Pn_2$) (or (Fe$_2$$Ch_2$)) blocks and play the main role in superconductivity, whereas the intermediate atomic sheets or oxide blocks serve as "charge reservoirs", see [2-9].

Very recently (R.J. Cava, et al., *arXiv*:1106.02111 (preprint, 10 June, 2011)), superconductivity with $T_C \sim 25$K was reported for the tetragonal phase Ca$_{10}$(Pt$_4$As$_8$)(Fe$_2$As$_2$)$_5$ forming in the quaternary Ca-Pt-Fe-As system [17].

Let us note that in the last period two additional reports became accessible (Dirk Johrendt et al., *arXiv*:1107.5320 (preprint, 26 July 2011), and Minoru Nohara, et al., *arXiv*:1108.0029 (preprint, 02 August 2011)), where the related Ca-PtFe-As systems are examined and where the superconductivity with a transition temperature up to $T_C \sim 38$K was reported [18,19].

One of the most intriguing features of these novel materials [17-19] is the presence of (Fe$_2$As$_2$) blocks, which are typical of the aforementioned family of iron-pnictide SCs, together with oxygen-free blocks (Pt$_n$As$_m$). Similar Pt-As blocks are known as the building blocks of recently discovered iron-free 122-like SrPt$_2$As$_2$ SC with the maximal $T_C \sim 5.2$K [20-22].

Based on the Zintl's chemical concept of ion electron counting, the authors [17] proposed that (Pt$_4$As$_8$) and (Fe$_2$As$_2$) blocks in the Ca$_{10}$(Pt$_4$As$_8$)(Fe$_2$As$_2$)$_5$ phase are metallic-like (*i.e.* both blocks will give appreciable contributions to the density of states at $E_F$), leading to enhanced inter-block coupling and thus to enhanced transition temperature of this system. It is declared also that further search of iron-pnictide phases with additional metallic-like blocks can provide an interesting platform for further search for novel high-$T_C$ superconducting materials.

One of central issues in realizing the properties of these new Ca-Pt-Fe-As materials are the data about the peculiarities of the electronic structure, chemical bonding and the distributions of atomic charges in $Ca_{10}(Pt_4As_8)(Fe_2As_2)_5$ which could confirm (or to deny) the mentioned speculations [17].

In order to get a detailed insight into the electronic properties and inter-atomic bonding picture for the newly discovered phase $Ca_{10}(Pt_4As_8)(Fe_2As_2)_5$, in the present work we performed first-principles calculations (by means of the full-potential linear augmented plane wave (FLAPW) method) of the optimized structural parameters, electronic bands, total and partial densities of states, and evaluated the peculiarities of inter-atomic interactions for this system, which were analyzed in comparison with the related layered superconducting materials.

## 2. Computational details

The examined phase $Ca_{10}(Pt_4As_8)(Fe_2As_2)_5$ crystallizes in a tetragonal structure (s.g. *P*4/*n*; # 85), which was classified [17] as a derivative of the $SrZnSb_2$ structure type. This structure can be schematically described as a sequence of 2D-like $(Pt_4As_8)$ and $(Fe_2As_2)_5$ blocks separated by Ca sheets; in turn, platinum-arsenide blocks $(Pt_4As_8)$ are formed by corner-sharing {$PtAs_4$} squares, whereas iron-arsenide blocks consist of {$FeAs_4$} tetrahedrons. In both cases $(Pt_4As_8)$ and $(Fe_2As_2)_5$ blocks contain a set of non-equivalent types of Fe, Pt, and As atoms, see Table 1, where the atomic positions are summarized. More detailed structural data for $Ca_{10}(Pt_4As_8)(Fe_2As_2)_5$ are discussed in Ref. [17].

Our band-structure calculations were carried out by means of the FLAPW method implemented in the WIEN2k suite of programs [23]. The generalized gradient correction (GGA) to exchange-correlation potential in the PBE form [24] was used. The plane-wave expansion was taken to $R_{MT} \times K_{MAX}$ equal to 8, and the *k* sampling with 8×8×7 *k*-points in the Brillouin zone was used. The calculations were performed with full structural optimization. The self-consistent calculations were considered to be converged when the difference in the total energy of the crystal did not exceed 0.1 mRy and the difference in the total electronic charge did not exceed 0.001 *e* as calculated at consecutive steps. The hybridization effects were analyzed using the densities of states (DOSs), which were obtained by the modified tetrahedron method [25]. The ionic bonding was considered using Bader's [26] analysis.

# 3. Results and discussion
## 3.1. Structural data

The calculated equilibrium atomic positions are presented in Table 1 and are in reasonable agreement with the available experiments [17]. The calculated lattice parameters $a = b = 8.794$ Å, and $c = 10.180$ Å also agree with experimental data $a = b = 8.7257$ Å, $c = 10.4243$ Å [17]. Some divergences are due, on the one hand, to the well-known peculiarities in estimations of the lattice parameters within LDA-GGA based calculation methods and, on the other hand, may be related to the quality of the examined $Ca_{10}(Pt_4As_8)(Fe_2As_2)_5$ single crystal, which contains approximately 9% of Ru and 7% of Pt in Fe sites [17].

## 3.2. Band structure and density of states

The electronic band structure for $Ca_{10}(Pt_4As_8)(Fe_2As_2)_5$ along the symmetry lines of the Brillouin zone (BZ) is shown in Fig. 1 as calculated for obtained equilibrium lattice parameters and atomic positions. The total and partial (atomic-resolved *l*-projected) densities of states (DOSs) were also calculated and are illustrated in Fig. 2.

It is seen that the electronic bands may be divided into two main groups, where the lowest group lying in the energy range from -13.2 eV to -9.8 eV below the Fermi level ($E_F$) arises mainly from As 4*s* states and is separated from the near-Fermi valence bands by a wide gap. This second group of electronic bands is located in the energy range from -6.7 eV to $E_F$ and is formed by the main contributions from the states of both $(Pt_4As_8)$ and $(Fe_2As_2)_5$ blocks. Let us note that the contributions from the valence states of Ca to the occupied bands are very small, *i.e.* in $Ca_{10}(Pt_4As_8)(Fe_2As_2)_5$ these atoms are in the form of cations. In turn, this means that the calcium sheets and the $(Pt_4As_8)$ and $(Fe_2As_2)_5$ blocks are linked mainly by ionic interactions, see also below.

The total DOS for $Ca_{10}(Pt_4As_8)(Fe_2As_2)_5$ demonstrates quite a complicated multi-peak structure which includes six main peaks A - F, Fig. 2. The lowest peak A arises mainly from As 4*p* states of $(Pt_4As_8)$ blocks, but the subsequent peaks (B-F) are of mixed types. So, peaks B and C contain hybridized Pt 5*d* - As 4*p* states of $(Pt_4As_8)$ blocks which are responsible for the covalent Pt-As bonds inside these blocks, together with the lowest As 4*p* states of $(Fe_2As_2)_5$ blocks. The intense peak D, along with Pt 5*d* - As 4*p* states of $(Pt_4As_8)$ blocks, includes hybridized Fe 3*d* - As 4*p* states of $(Fe_2As_2)_5$ blocks which participate in covalent Fe-As bonds inside these blocks. The near-Fermi peaks E and F are originated mainly from Fe 3*d* electronic states. These states contribute to Fe-Fe bonds and are responsible for the metallic-like character of this system. Note that similar results have been obtained recently within LMTO calculations of DOSs for tetragonal $Ca_{10}(Pt_4As_8)(Fe_2As_2)_5$ as well as for triclinic $Ca_{10}(Pt_3As_8)(Fe_2As_2)_5$ systems [18].

Let us underline the considerable differences in the electronic structures of $(Pt_4As_8)$ *versus* $(Fe_2As_2)_5$ blocks. So, the $(Fe_2As_2)_5$ blocks are metallic-like, and the electronic bands in the window around the Fermi level are formed mainly by the Fe $3d$ states. Here, the As $4p$ states are located in a narrow enough interval from 5.6 eV to -2.3 eV.

On the contrary, for the $(Pt_4As_8)$ blocks, the formation of a pseudo-gap around the Fermi level with very low densities of states at the Fermi level ($N(E_F)$) was established; besides, the contributions to $N(E_F)$ come both from the Pt $5d$ and As $4p$ states, Table 2. Thus, unlike the assumption [17], our data show that $(Pt_4As_8)$ blocks will behave as semi-metals. Besides, the As $4p$ states of $(Pt_4As_8)$ blocks occupy a very wide energy interval coinciding with the width of the common valence band, ~ 6.6 eV. This can be explained by participation of the As $4p$ states in the formation of Pt-As and As-As covalent bonds, see also below.

As electrons near the Fermi level are involved in the formation of the superconducting state, let us discuss these states in more detail, see Fig. 1 and Table 2, where the values of $N(E_F)$ for related phases $SrFe_2As_2$ and $SrPt_2As_2$ are also presented.

We see that the near-Fermi bands demonstrate a unique complicated "mixed" character: simultaneously with quasi-flat bands with low $E(k)$ dispersion along Γ-Z, a series of high-dispersive bands intersects the Fermi level between Z and R, M and A points and in the A-Z direction. Simultaneously, the calculated DOSs at the Fermi level demonstrate that the main contribution to $N(E_F)$ is from the Fe $3d$ states; the contributions from other orbitals to $N(E_F)$ are very small. Thus, conduction in $Ca_{10}(Pt_4As_8)(Fe_2As_2)_5$ is expected to be anisotropic and happening mainly in $(Fe_2As_2)_5$ blocks.

*3.3. Chemical bonding*

Let us discuss the peculiarities of inter-atomic bonding for the $Ca_{10}(Pt_4As_8)(Fe_2As_2)_5$ phase, which seem to be very important for the understanding of inter-atomic and inter-blocks charge transfer.

Indeed, assignment of the usual oxidation numbers of atoms $Ca^{2+}$, $Fe^{2+}$, and $As^{3-}$ for the iron arsenide blocks and $Pt^{2+}$ and $As^{2-}$ for the platinum arsenide blocks (where the formation of $As_2^{4-}$ dimers was assumed) [17] immediately gives an unbalanced ionic formula $Ca_{10}^{20+}(Pt_4As_8)^{8-}(Fe_2As_2)_5^{10-}$; this implies infringement of electroneutrality for this crystal. The unbalanced valence state $(Ca_{10}^{20+}(Pt_4As_8)^{16-}(Fe_2As_2)_5^{10-})$ also takes place if we use the usual oxidation number (-3) for the As atoms in $(Pt_4As_8)$.

Thus, the actual charge states for $Ca_{10}(Pt_4As_8)(Fe_2As_2)_5$ are far from the ideal ionic picture. To estimate the atomic charges in this crystal, which strictly obey the electroneutrality rule, the Bader's scheme [24] was applied. In this approach, each atom of a crystal is surrounded by an effective surface that runs through minima of the charge density, and the total charge of an atom (so-called Bader's charge, $Q^B$) is determined by integration within this region. The values of $Q^B$ are presented in Table 1. Then the effective atomic charges ($\Delta Q$) are estimated as $\Delta Q = Q^B - Q^i$, where $Q^i$ are ionic charges in the assumption of the aforementioned oxidation numbers: $Ca^{2+}$, $Pt^{2+}$, $Fe^{2+}$, and $As^{3-}$ - for the iron arsenide blocks. In this way, the calculated $\Delta Q$ (Table 1) provide the charge states $Ca_{10}^{14.144+}$ and $(Fe_2As_2)_5^{1.892-}$. This yields a charge-balanced platinum arsenide block $(Pt_4As_8)^{12.252-}$. Hence, using the calculated values of $\Delta Q$ for Pt1,2 atoms (Table 1), we can estimate the effective atomic charges of As3 atoms inside $(Pt_4As_8)$ blocks: $\Delta Q = -2.106$.

The above analysis allows us to make the following conclusions:

(i). The ionic state of As3 atoms inside $(Pt_4As_8)$ blocks (-2.434) adopts an "intermediate" type between "ideal" oxidation numbers $As^{3-}$ and $As^{2-}$, which are characteristic of arsenic ions inside $(Fe_2As_2)$ blocks and in assumed dimers $As_2^{4-}$, respectively, and is determined by features of As-As bonding inside $(Pt_4As_8)$ blocks, see below;

(ii). Our analysis indicates the presence of appreciable charge anisotropy in $Ca_{10}(Pt_4As_8)(Fe_2As_2)_5$ crystal both between structurally non-equivalent atoms (for example Fe1-Fe3, Pt1-Pt2 *etc.*, see Table 1) and between the main building blocks: $(Pt_4As_8)$ *versus* $(Fe_2As_2)_5$.

(iii). If the difference in the Bader's charges, $Q^B$, between non-equivalent Fe1,2, Ca1,2, and Pt1,2 atoms does not exceed ~ 0.13 *e* (Table 1), the maximum charge anisotropy ($\delta Q^B \sim 0.60$ *e*) arises between the As1,2 and As3 atoms, which are placed inside $(Pt_4As_8)$ and $(Fe_2As_2)_5$ blocks, respectively. This situation can be explained taking into account the formation by As3 atoms of direct Pt-As and As-As bonds inside $(Pt_4As_8)$ blocks, which are well visible in Fig. 3.

(iv). It is useful to compare the effective atomic charges, $\Delta Q$, for the examined $Ca_{10}(Pt_4As_8)(Fe_2As_2)_5$ crystal with the same values calculated for some related phases which include similar iron-arsenide blocks or platinum-arsenide blocks. So, for LaOFeAs, $SrFe_2As_2$, and LiFeAs, $\Delta Q(Fe)$ are +1.722, +1.912, and +1.821, whereas $\Delta Q(As)$ are -2.113, -2.241, and -1.995, respectively [27]. Thus, the effective atomic charges for $(Fe_2As_2)_5$ blocks in $Ca_{10}(Pt_4As_8)(Fe_2As_2)_5$ are comparable with the same values for earlier known related superconducting materials. On the other hand, the effective atomic charges for $SrPt_2As_2$ ($\Delta Q(Pt) = +2.6$ and $\Delta Q(As) = -3.0$ [22]) are higher than those for $(Pt_4As_8)$ blocks in the examined phase owing to formation of the aforementioned As-As covalent bonds.

(v). Finally, the inter-blocks charge transfer occurs from the electropositive calcium ions to $(Pt_4As_8)$ and $(Fe_2As_2)_5$ blocks; and here the charge transfer $Ca_{10} \rightarrow (Pt_4As_8)$ (about 12 *e*) is much greater than the charge transfer $Ca_{10} \rightarrow (Fe_2As_2)_5$, which is about 2 *e*. To illustrate this, in Fig. 4 we plotted the iso-electronic surface for $Ca_{10}(Pt_4As_8)(Fe_2As_2)_5$. Thus, in contrast to the majority of known superconducting iron-containing materials [2-9], the new phase $Ca_{10}(Pt_4As_8)(Fe_2As_2)_5$ includes two negatively charged blocks, where the charge of superconducting $[Fe_2As_2]_5$ blocks is much smaller than for the platinum arsenide blocks. On the other hand, the charge state of the iron arsenide blocks in $Ca_{10}(Pt_4As_8)(Fe_2As_2)_5$ (containing about two additional electrons) is comparable with the same for the iron arsenide blocks in other superconducting iron-containing materials, see [2-9].

## 4. Conclusions

In conclusion, we presented a first principles study of the tetragonal crystal $Ca_{10}(Pt_4As_8)(Fe_2As_2)_5$ which was very recently proposed [17-19] as a parent phase for the new family of high-$T_C$ iron-pnictide superconductors with intermediate skutterudite-like blocks. Our analysis comprises the optimized structural parameters, electronic bands, total and partial densities of states, and inter-atomic bonding picture for this material - in comparison with those for the related layered superconducting iron-based systems.

As a result, we found that like the known iron-based systems, $Ca_{10}(Pt_4As_8)(Fe_2As_2)_5$ is metallic-like, and the electronic bands in the window around the Fermi level are formed mainly by the Fe $3d$ states of $(Fe_2As_2)_5$ blocks. Thus, conduction in $Ca_{10}(Pt_4As_8)(Fe_2As_2)_5$ is expected to be anisotropic and happening mainly in $(Fe_2As_2)_5$ blocks.

On the other hand, unlike the assumption [17], we found that the $(Pt_4As_8)$ blocks will behave as semi-metals with very low densities of states at the Fermi level, and the contributions to $N(E_F)$ come both from Pt $5d$ and As $4p$ states. Besides, for $Ca_{10}(Pt_4As_8)(Fe_2As_2)_5$ the near-Fermi bands adopt a unique complicated "mixed" character: simultaneously with quasi-flat bands with low $E(k)$ dispersion, a series of high-dispersive bands which intersect the Fermi level, was found.

The picture of chemical bonding in $Ca_{10}(Pt_4As_8)(Fe_2As_2)_5$ is very complicated: it can be described as an anisotropic mixture of covalent, metallic, and ionic contributions and includes a set of various inter-atomic and inter-block interactions. Namely, inside $(Fe_2As_2)_5$ blocks covalent Fe-As and metallic-like Fe-Fe bonds take place, whereas inside $(Pt_4As_8)$ blocks a system of covalent Pt-As and As-As bonds emerges. Besides, inside these blocks the inter-atomic ionic interactions occur owing to charge transfer Fe → As and Pt → As.

Finally, the inter-blocks charge transfer occurs from the electropositive Ca ions to $(Pt_4As_8)$ and $(Fe_2As_2)_5$ blocks. It is important that the charge transfer $Ca_{10}$ → $(Pt_4As_8)$ is much greater than the transfer $Ca_{10}$ → $(Fe_2As_2)_5$, *i.e.* unlike the majority of known superconducting iron-containing materials [2-9], the new phase $Ca_{10}(Pt_4As_8)(Fe_2As_2)_5$ includes two negatively charged blocks, where the charge of conducting $(Fe_2As_2)_5$ blocks is much smaller than for the Pt-As blocks.

## Acknowledgments

The authors acknowledge the support from the RFBR (grants No. 09-03-00946 and No. 10-03-96008).

**Table 1**
Equilibrium atomic positions, Bader's charges, ($Q^B$, in $e$), and atomic effective charges ($\Delta Q$, in $e$) for $Ca_{10}(Pt_4As_8)(Fe_2As_2)_5$ as obtained within FLAPW-GGA calculations.

| Atom (position) | | Atomic coordinates | | | Atomic charges | |
|---|---|---|---|---|---|---|
| | | x/a | y/b | z/c | $Q^B$ | $\Delta Q$ |
| Ca1 | 2c | ¼ | ¼ | 0.7650 (0.7736) | 6.584 | +1.416 |
| Ca2 | 8g | 0.3449 (0.3454) | 0.9532 (0.9520) | 0.2505 (0.2417) | 6.586 | +1.414 |
| Fe1 | 2b | ¾ | ¼ | ½ | 7.856 | +1.856 |
| Fe2 | 8g | 0.4496 (0.4570) | 0.1501 (0.1550) | 0.4983 (0.5020) | 7.851 | +1.851 |
| Pt1 | 2c | ¼ | ¼ | 0.0780 (0.0692) | 10.784 | +1.216 |
| Pt2 | 2a | ¼ | ¾ | 0 | 10.921 | +1.079 |
| As1 | 8g | 0.5485 (0.54824) | 0.3493 (0.3499) | 0.6228 (0.6364) | 5.967 | -2.033 |
| As2 | 2c | ¼ | ¼ | 0.3749 (0.3603) | 5.926 | -2.074 |
| As3 | 8g | 0.5105 (0.5095) | 0.8586 (0.8590) | 0.0206 (0.0169) | 5.328 | -2.106 |

* the available experimental data (Ref. [17]) are given in parenthesis

**Table 2**
Total and partial (in states/eV·f.u.) densities of states at the Fermi level for $Ca_{10}(Pt_4As_8)(Fe_2As_2)_5$ as obtained within FLAPW-GGA calculations in comparison with the same data for related layered phases $SrFe_2As_2$ and $SrPt_2As_2$ [22].

| DOSs | $Ca_{10}(Pt_4As_8)(Fe_2As_2)_5$ | $SrFe_2As_2$ | $SrPt_2As_2$ |
|---|---|---|---|
| Total | 28.54 | 3.74 | 2.55 |
| Fe1 3d | 4.35 (2.18) * | 1.90 | - |
| Fe2 3d | 16.74 (2.09) | - | - |
| Pt1 5d | 0.34 (0.17) | - | 0.69 ** |
| Pt2 5d | 0.15 (0.08) | - | 0.30 ** |
| As1 4p | 0.31 (0.04) | 0.12 | 0.11 ** |
| As2 4p | 0.10 (0.05) | - | 0.16 ** |
| As3 4p | 0.24 (0.03) | - | - |

* the atomic-resolved $l$-projected DOSs per one atom are given in parenthesis
** for non-equivalent Pt1,2 and As1,2 atoms in $CaBe_2Ge_2$-type $SrPt_2As_2$; see Ref. [22].

**FIGURES**

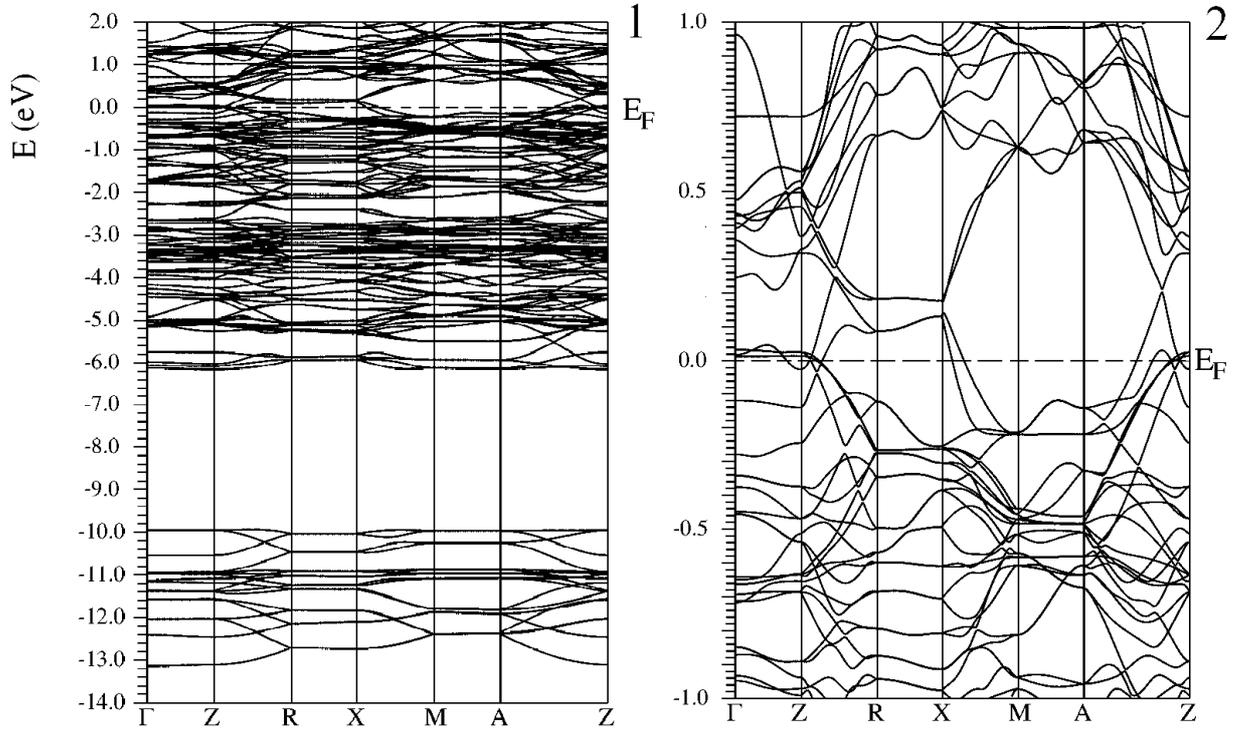

**Fig. 1.** Valence (1) and near-Fermi (2) electronic bands for $Ca_{10}(Pt_4As_8)(Fe_2As_2)_5$. The Fermi level $E_F = 0$ eV.

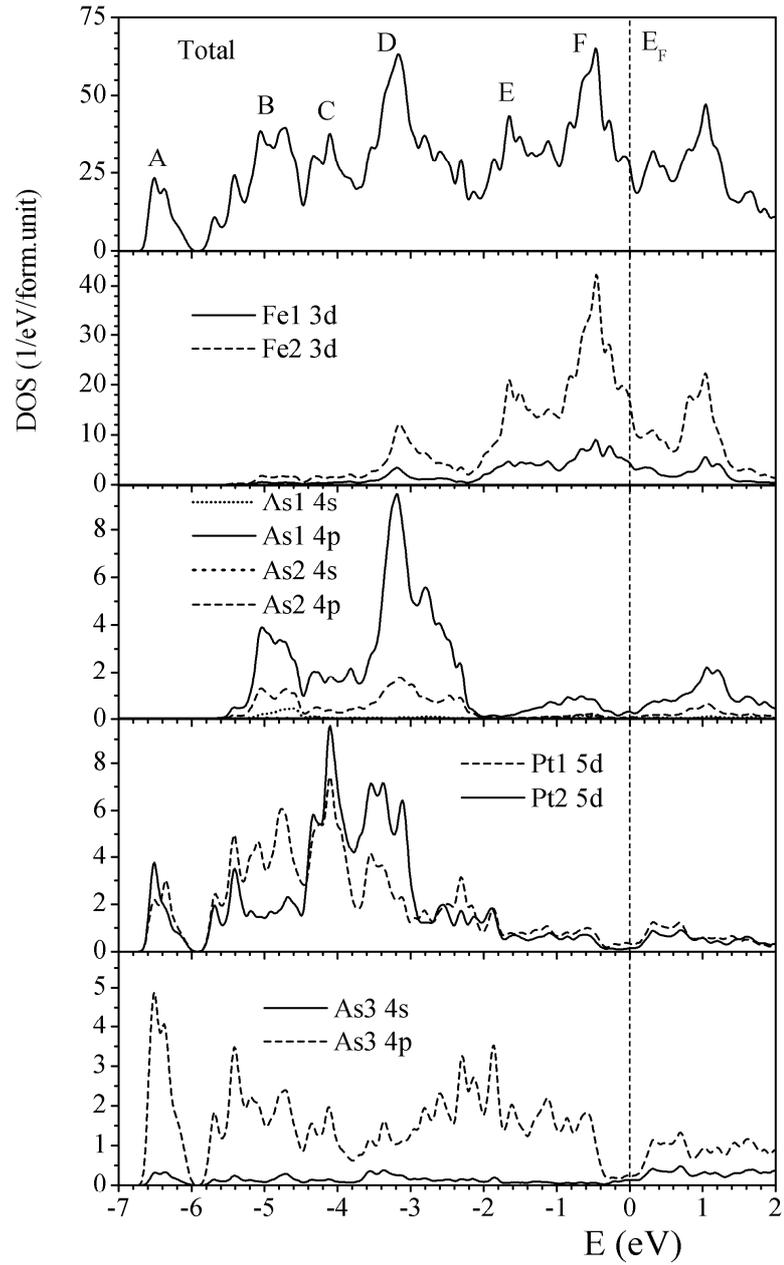

**Fig. 2.** Total (*upper panel*) and partial (*bottom panels*) densities of states for $Ca_{10}(Pt_4As_8)(Fe_2As_2)_5$. The Fermi level $E_F = 0$ eV.

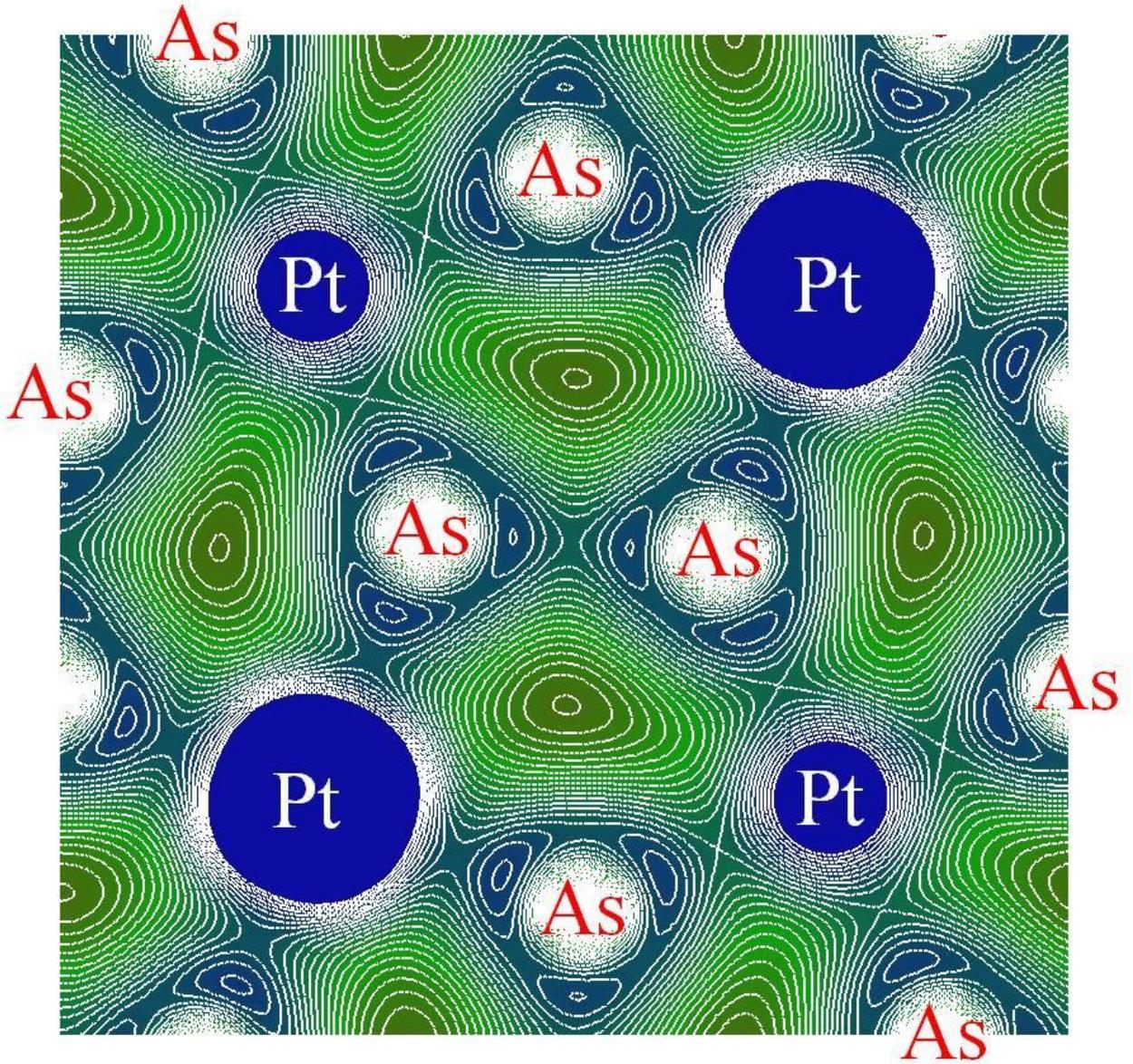

**Fig. 3.** (*Color online*) Charge density map in ($z_{As3}$00) plane for $Ca_{10}(Pt_4As_8)(Fe_2As_2)_5$ phase, which illustrates the formation of directional covalent As-As bonds inside ($Pt_4As_8$) blocks, *see the text*.

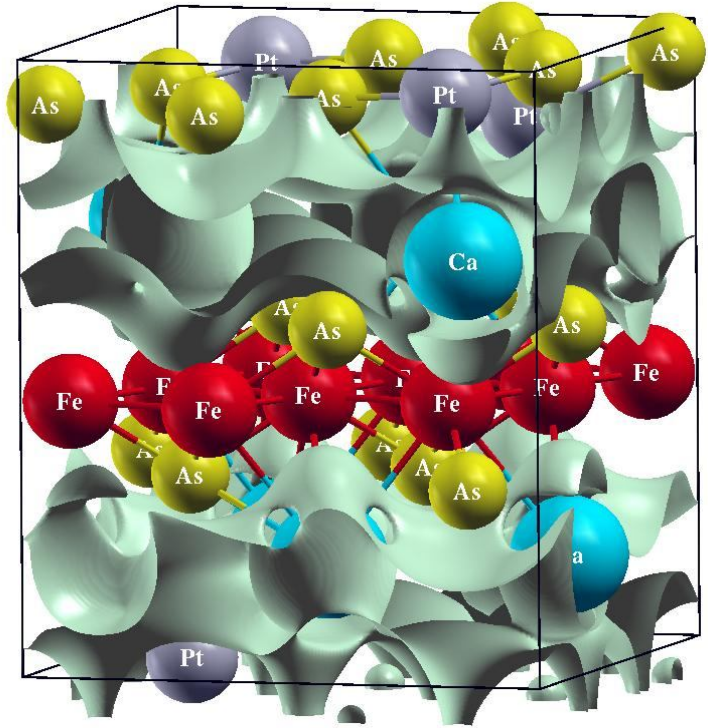

**Fig. 4.** (*Color online*). Charge density iso-surface (ρ = 0.11 e/Å$^3$) for $Ca_{10}(Pt_4As_8)(Fe_2As_2)_5$ which illustrates the maximal charge transfer in direction $Ca_{10} \rightarrow (Pt_4As_8)$, *see the text*.